\documentclass[twocolumn,floatfix,superscriptaddress,a4paper,showpacs,showkeys,nofootinbib,reprint,prc,noeprint]{revtex4-1}
\usepackage{epsfig}
\usepackage{latexsym}
\usepackage[colorlinks=true,linktocpage=true,linkcolor=blue,citecolor=blue,allcolors=blue]{hyperref}
\usepackage{url}
\usepackage[utf8]{inputenc}
\usepackage{enumerate}
\usepackage{color}
\usepackage{xcolor}

\usepackage{xfrac}

\usepackage{amsmath}
\usepackage{amssymb}

\usepackage[english]{babel}
\usepackage{hyperref}
\usepackage{url}
\usepackage{needspace}
\usepackage{float}
\hypersetup{
   colorlinks=true, 
  linktoc=all,     
  linkcolor=blue,  
}

\newcommand{\dd}{{\rm d}}

\newcommand{\cR}[2]{c_{#1}^{#2}}

\begin{document}

 \title{Determining the Duration of the Hadronic Stage at RHIC-BES Energies via Resonance Suppression Using a Full Set of Rate Equations}

\author{Tim Neidig}
\affiliation{Institut f\"{u}r Theoretische Physik, Goethe Universit\"{a}t Frankfurt, Max-von-Laue-Str. 1, D-60438 Frankfurt am Main, Germany}
\affiliation{Helmholtz Research Academy Hesse for FAIR (HFHF), GSI Helmholtzzentrum f\"ur Schwerionenforschung GmbH, Campus Frankfurt, Max-von-Laue-Str. 12, 60438 Frankfurt am Main, Germany}

\author{Apiwit Kittiratpattana}
\affiliation{Institut f\"{u}r Theoretische Physik, Goethe Universit\"{a}t Frankfurt, Max-von-Laue-Str. 1, D-60438 Frankfurt am Main, Germany}
\affiliation{Center of Excellence in High Energy Physics and Astrophysics, School of Physics, Suranaree University of Technology, University Avenue 111, Nakhon Ratchasima 30000, Thailand}

\author{Tom Reichert}
\affiliation{Institut f\"{u}r Theoretische Physik, Goethe Universit\"{a}t Frankfurt, Max-von-Laue-Str. 1, D-60438 Frankfurt am Main, Germany}
\affiliation{Helmholtz Research Academy Hesse for FAIR (HFHF), GSI Helmholtzzentrum f\"ur Schwerionenforschung GmbH, Campus Frankfurt, Max-von-Laue-Str. 12, 60438 Frankfurt am Main, Germany}

\author{Amine Chabane}
\affiliation{Institut f\"{u}r Theoretische Physik, Goethe Universit\"{a}t Frankfurt, Max-von-Laue-Str. 1, D-60438 Frankfurt am Main, Germany}
\affiliation{Helmholtz Research Academy Hesse for FAIR (HFHF), GSI Helmholtzzentrum f\"ur Schwerionenforschung GmbH, Campus Frankfurt, Max-von-Laue-Str. 12, 60438 Frankfurt am Main, Germany}

\author{Carsten Greiner}
\affiliation{Institut f\"{u}r Theoretische Physik, Goethe Universit\"{a}t Frankfurt, Max-von-Laue-Str. 1, D-60438 Frankfurt am Main, Germany}
\affiliation{Helmholtz Research Academy Hesse for FAIR (HFHF), GSI Helmholtzzentrum f\"ur Schwerionenforschung GmbH, Campus Frankfurt, Max-von-Laue-Str. 12, 60438 Frankfurt am Main, Germany}

\author{Marcus Bleicher}
\affiliation{Institut f\"{u}r Theoretische Physik, Goethe Universit\"{a}t Frankfurt, Max-von-Laue-Str. 1, D-60438 Frankfurt am Main, Germany}
\affiliation{Helmholtz Research Academy Hesse for FAIR (HFHF), GSI Helmholtzzentrum f\"ur Schwerionenforschung GmbH, Campus Frankfurt, Max-von-Laue-Str. 12, 60438 Frankfurt am Main, Germany}

\date{\today}

\begin{abstract}
We present realistic estimates for the duration of the hadronic stage in central Au+Au reactions in the RHIC-BES energy regime. To this aim, we employ a full set of coupled rate equations to describe the time evolution of the system from chemical to kinetic freeze-out. Combined with the recently measured data by the STAR collaboration on $K^*/K$ ratios, we show that the previous estimates substantially underestimated the duration of this stage due to the omission of the regeneration of hadron resonances. We provide an improved relation between the $K^*/K$ ratio at chemical and kinetic freeze-out and the life time of the hadronic phase. The calculated improved life times are now in line with estimates from other methods and are relevant for the NA61 and STAR collaborations and for upcoming experiments at the FAIR facility.
\end{abstract}

\maketitle

\section{Introduction}
Heavy-ion collisions are today's main tool to explore the properties of hot and dense QCD matter \cite{Sorensen:2023zkk}. A general picture of the evolution of a heavy-ion collision consists of an initial state, followed by an expansion of the fireball (either composed of hadrons or quarks and gluons) until the inelastic interactions cease. This point in time is often referred to as the chemical freeze-out. After the chemical freeze-out further (mainly elastic) collisions happen until the system breaks-up (called the kinetic freeze-out) and the hadrons escape to the detector \cite{Bleicher:2002dm,Andronic:2005yp}. Measuring the duration of a heavy-ion collision or the durations of the individual stages is still a challenge. A well established tool to obtain information on the time scales is Hanbury-Brown-Twiss (HBT) interferometry or nowadays called Femtoscopy \cite{HanburyBrown:1956bqd,Lisa:2005dd}. 

An alternative way to study the duration of the hadronic stage is based on the suppression of hadron resonances \cite{Torrieri:2001ue,Rafelski:2001hp,Bleicher:2002dm, LeRoux:2021adw,Knospe:2021jgt,Knospe:2015nva,Ilner:2017tab,Oliinychenko:2021enj,Chabane:2024crn}. The main idea is that hadron resonances, e.g. the mesonic $K^*$, $\rho$, $\phi$ or the baryonic $\Delta$, $\Lambda^*$, $\Sigma^*$, etc. are created at the chemical freeze-out, subsequently they decay and the experiment reconstructs them from their hadronic decay products. The life time of the hadronic stage can then be estimated from the relative suppression of the resonance over groundstate ratio, because the decay products of a resonance that decays rescatter after the chemical freeze-out until the end of the kinetic stage, respectively until kinetic freeze-out. The longer the life time of the hadronic stage, the more suppression of the initial resonance yield is observed because its decay daughters rescatter more leading to increased signal loss. This suppression has been measured and used to extract life times from GSI, to RHIC-BES and up to the LHC \cite{STAR:2004bgh,Markert:2005jv,STAR:2008twt,HADES:2013sfy,Knospe:2015rja,ALICE:2018ewo,ALICE:2022zuc,ALICE:2023ifn}, and the NA61 collaboration is further currently investigating the $K^*/K$ ratio in smaller systems \cite{Kozlowski:2024cjw}.
Also an alternative scenario has been put forward that does not assume a signal loss due to rescattering, but attributes the suppression to the cooling of the hadronic system in partial chemical equilibrium which suppresses the resonance yields via a substantially lowered freeze-out temperature, see e.g. \cite{Motornenko:2019jha,Neidig:2021bal,LeRoux:2021adw}. This also allows to estimate the duration of the hadronic stage and is the approach we are going to follow in this work.

The idea of relating the life time of the hadronic phase to the suppression of resonances was put forward by the STAR collaboration \cite{STAR:2002npn,STAR:2022sir} to obtain a lower bound for the duration of the hadronic rescattering phase in heavy-ion collisions. Following their approach leads to the relation
\begin{equation}
    \left( \frac{h^*}{h} \right)\bigg|_{\text{KFO}} = \left( \frac{h^*}{h} \right)\bigg|_{\text{CFO}} \times \exp\left( - \frac{\Delta t_{\text{hadronic}}}{\tau_{h^*}} \right).
\label{eq:star}    
\end{equation}
where $h$ is a groundstate hadron, $h^*$ is the corresponding resonance state, $\Delta t_{\text{hadronic}}$ is the life time of the hadronic rescattering phase, $\tau_{h^*} = 1/\Gamma^{h^{\star}}_{\text{vac}}$ is the (vacuum) life time of the $h^*$ hadronic resonance and ``CFO" and ``KFO" denote the chemical and kinetic freeze-out, respectively.

However, this picture is too simplified \cite{Chabane:2024crn} as it assumes that the interaction probability for the decay daughters \cite{Torrieri:2001ue,Torrieri:2001tg} is large and the gain (regeneration) and loss  rates can be neglected during the hadronic stage.

Here we overcome these limitations by employing a full set of kinetic equations to describe the resonance's time evolution: 
\begin{equation}
    N^{h^*}_\text{kin} - N^{h^*}_\text{chem} = (-\Gamma^{h^*} N^{h^*} - L_\text{coll} + G_\text{reg} ) \Delta t_\text{hadronic},
\end{equation}
where the time is integrated from the chemical freeze-out to the kinetic freeze-out (life time of the hadronic phase, $\Delta t_\text{hadronic} = t_\mathrm{kin} - t_\mathrm{chem}$). $\Gamma^{h^*}$ is the width of the $h^*$, $ L_\text{coll}$ is the loss rate due to collisions of the $h^*$ and $G_\text{reg}$ is a gain rate, e.g. $h+\pi \rightarrow h^*$. 

\section{Framework}
To model the time evolution of the resonance production and decay in the hadronic environment the framework of partial chemical equilibrium (PCE) \cite{Bebie:1991ij} to describe the hadronic environment after the chemical freeze-out is a well established method. 
Here, the total yield of stable hadrons is fixed by introducing non-equilibrium chemical potentials~\cite{Xu:2018jff,Vovchenko:2019aoz}, while the yields of resonances changes \cite{Motornenko:2019jha,Tomasik:2021jfd,Neidig:2021bal} and is being determined by the relative equilibrium of decays and regenerations. This means, we avoid the assumption of instantaneous equilibration of nuclear reactions, as well as resonance decays and regenerations. This is achieved by employing a set of coupled (reaction) rate equations describing the chemical composition of stable hadrons (pions, kaons, (anti-)nucleons, $\Lambda $s), as well as the hadronic resonances. Such a description of the
chemical evolution is well known in relativistic heavy-ion collisions (see e.g. for (multi-) strange baryons \cite{Koch:1986ud,Barz:1988md,Noronha-Hostler:2007fzh,Noronha-Hostler:2009wof}) and in astrophysics, see e.g. \cite{Kolb:1990vq}. 

Here we focus on the dynamics of the hadron resonances, i.e. their decay and formation by the reaction and employ the specific rate equations described in \cite{Neidig:2021bal}:
\begin{equation}
    h^* \rightleftharpoons X + h\ .
\label{eq:46}
\end{equation}
In the case of resonance decays, the change in the multiplicity of resonance $h^*$ with time $t$ can be expressed as
\begin{equation}
\label{eq:reso_rate}
  \frac{\dd N_{h^*}}{\dd t}
  = -\langle \Gamma_{h^* \rightarrow h + X} \rangle (N_{h^*}-\cR{h^*}{hX}N_h N_X)\,
\end{equation}
where $\langle \Gamma_{h^* \rightarrow h + X} \rangle$ can be interpreted as the (thermal averaged) decay rate of $h^*\to h+X$, but also as a scaled cross section for $X+h \to h^*$. The factor $\cR{h^*}{X h} = N_h^{*,\rm eq}/(N_X^{\rm eq}N_h^{\rm eq})$, given by the ratio of the multiplicities in equilibrium, $N_i^{\rm eq}$, is dictated by the detailed balance. Any $N_i^{\rm eq}$ is calculated at full thermal and chemical equilibrium.
The contributions of resonance $h^*$ decays and regenerations to the change rate of the multiplicities of the decay products, $\dd N_{X/h}/\dd t$, are given by the rhs. of Eq.~\eqref{eq:reso_rate} with the opposite sign.

\begin{align}
\label{eq:back_rate}
  \frac{\dd N_{h/X}}{\dd t}
  &= -\frac{\bigl\langle \sigma_{_{h+X \rightarrow h^*}} v_{\rm rel}\bigr\rangle}{V} (N_h N_X -\cR{hX}{h^*} N_{h^*})
\end{align}

Note, that because of the detailed balance condition, the cross section in (\ref{eq:back_rate}) and the decay width in (\ref{eq:reso_rate}) are related via,

\begin{align}
\label{eq:detailed_balance}
  \frac{\bigl\langle \sigma_{_{h+X \rightarrow h^*}} v_{\rm rel}\bigr\rangle}{V} \cR{hX}{h^*}
  &= \langle \Gamma_{h^* \rightarrow h + X} \rangle.
\end{align}

The decay rates employed are thermal-averaged values derived from experimentally measured cross sections or resonance decay widths sourced from Particle Data Tables~\cite{ParticleDataGroup:2020ssz}.

The multiplicity ratio of resonances over their ground state in heavy-ion collisions is calculated by including the expansion and cooling of the fireball in the hadronic phase. This involves a time dependence of the volume, $V(t)$, and the temperature, $T(t)$.
In the framework of the PCE, the temperature $T$ is obtained for a given (expanding) volume $V$ in such a way that the abundance of each stable hadron species (including the resonance contribution) is conserved. Furthermore, the total entropy of the system is conserved, as are the net baryon number and the net strangeness.
These relations provide a set of non-linear equations that can be solved for any $T \le T_{\rm ch}$ in order to obtain the volume, $V(T)$, and the chemical potential of each particle, $\mu_{i}(T)$ (i = S, B, N, $\pi$ ...). The initial conditions at $T_{\rm ch}$, namely $V(T_{\rm ch})$ and the baryon and strangeness chemical potential, are obtained from the standard parametrization

\begin{align}
    \mu_\mathrm{B,ch}(\sqrt{s_\mathrm{NN}}) &= \frac{1.308\,\mathrm{GeV}}{1 + 0.273 \, \mathrm{GeV}^{-1}\sqrt{s_\mathrm{NN}}}
    \label{eq:pbm_mu} 
\end{align}

\begin{align}
T_\mathrm{ch}(\mu_\mathrm{B,ch}) &= 0.166 \, \mathrm{GeV} - 0.139 \, \mathrm{GeV}^{-1} \, \mu_\mathrm{B,ch}^{2} \nonumber \\ 
&- 0.053 \, \mathrm{GeV}^{-3} \, \mu_\mathrm{B,ch}^{4} 
\label{eq:pbm_T} 
\end{align}

taken from \cite{Cleymans:2006qe} and we have parametrized the volume at chemical freeze-out (based on the data summarized in \cite{Andronic:2009qf}) as 

\begin{align}
V_\mathrm{ch}(x) = (2351.7 x^2 -10009.5 x + 13878.4) x^{-1.36} \; \mathrm{fm}^3 . 
\label{eq:Tom_V} 
\end{align}

where $x \equiv \ln(\sqrt{s_\mathrm{NN}}/\mathrm{GeV})$.
The initial strangeness chemical potential is obtained by forcing the total strangeness to zero and solving the equation,

\begin{align}
0 &= V_\mathrm{ch} \sum_{j \in \mathrm{particles}} s_j \, n_{j}(T_\mathrm{ch}, \mu_\mathrm{B,ch} , \mu_\mathrm{S,ch}),
\label{eq:stange_constrain} 
\end{align}

with $s_j$ denoting the strangeness of the corresponding particle species and $n_j$ their density.

The time dependence of the volume is given by the parametrization that incorporates longitudinal and transverse expansion \cite{Pan:2014caa},
\begin{equation}
    V(t)= V_{\rm ch} \frac{t}{t_{\rm ch}} \frac{t_{\perp}^{2}+t^{2}}{t_{\perp}^{2}+t_{\rm ch}^{2}}
\label{eq:V}
\end{equation}
with $t_{\perp}=6.5$~fm/c. We fix $t_{\rm ch}$, based on previous UrQMD calculations \cite{Reichert:2020yhx}. The centrality dependence of the volume $V_{\rm ch}$ is obtained by scaling the volume with the number of charged particles. The value of $t_{\perp}$ is kept constant, as it has only minor influence on the final results.
In the present work, all calculations assume an initial temperature and chemical potential for the chemical freeze-out as provided by Eqs. \eqref{eq:pbm_mu}, \eqref{eq:pbm_T} for each individual collision energy.

The network contains 23 rate equations, given by the number of stable and unstable particles and antiparticles. For further details, we refer the reader to \cite{Neidig:2021bal}.

\begin{figure} [t!]
    \centering
    \includegraphics[width=\columnwidth]{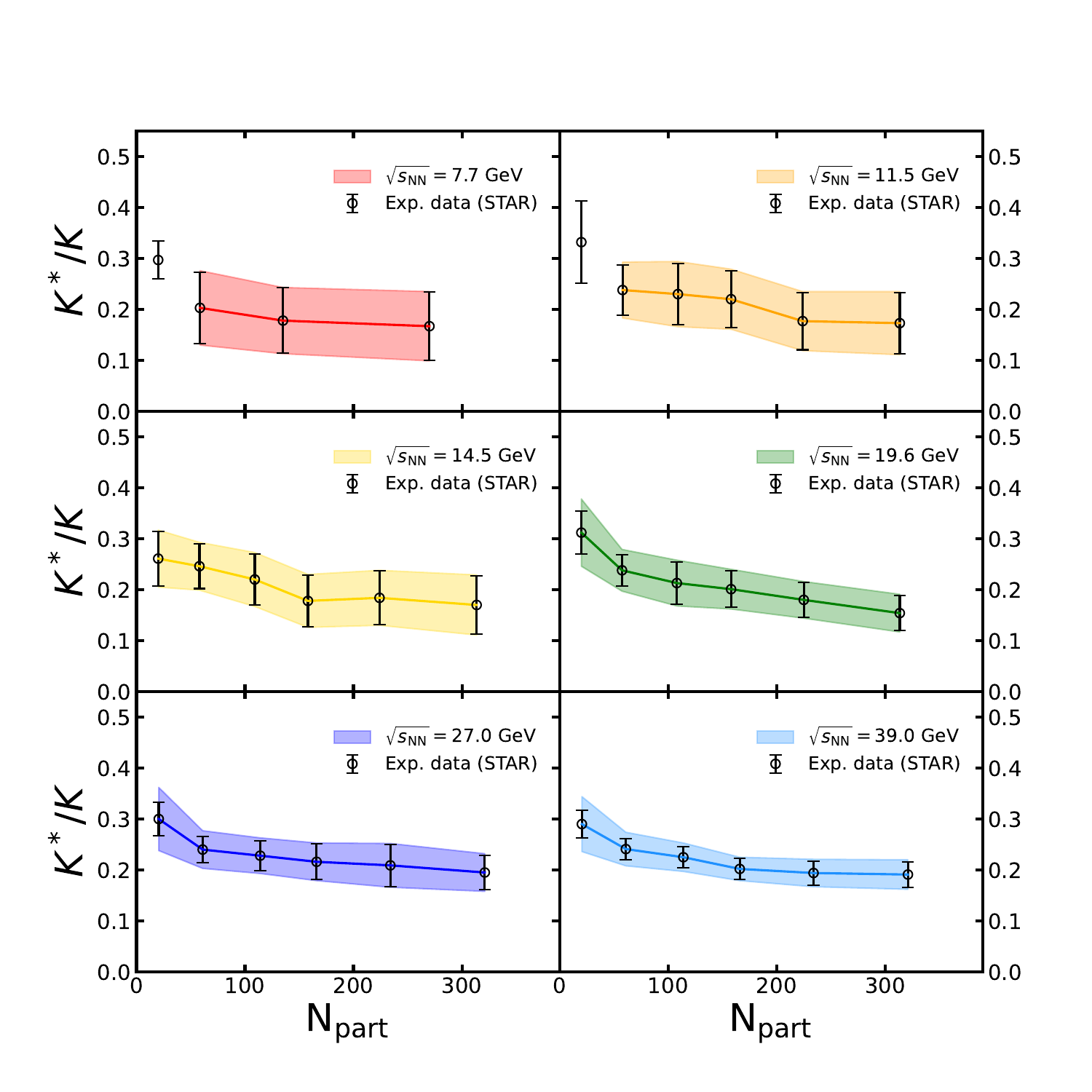}
    \caption{[Color online] $K^*/K$ ratio in Au+Au reactions as a function of centrality in the energy range from $\sqrt{s_\mathrm{NN}}=7.7$ GeV to $\sqrt{s_\mathrm{NN}}=39$ GeV (shown from upper left to bottom right). Full colored symbols with band denote the rate equation network calculations, while the open black symbols show the experimental data from the STAR collaboration \cite{STAR:2022sir}.}
    \label{fig:Multiplicity_ratios_edep}
\end{figure}

\begin{figure} [t!]
    \centering
    \includegraphics[width=\columnwidth]{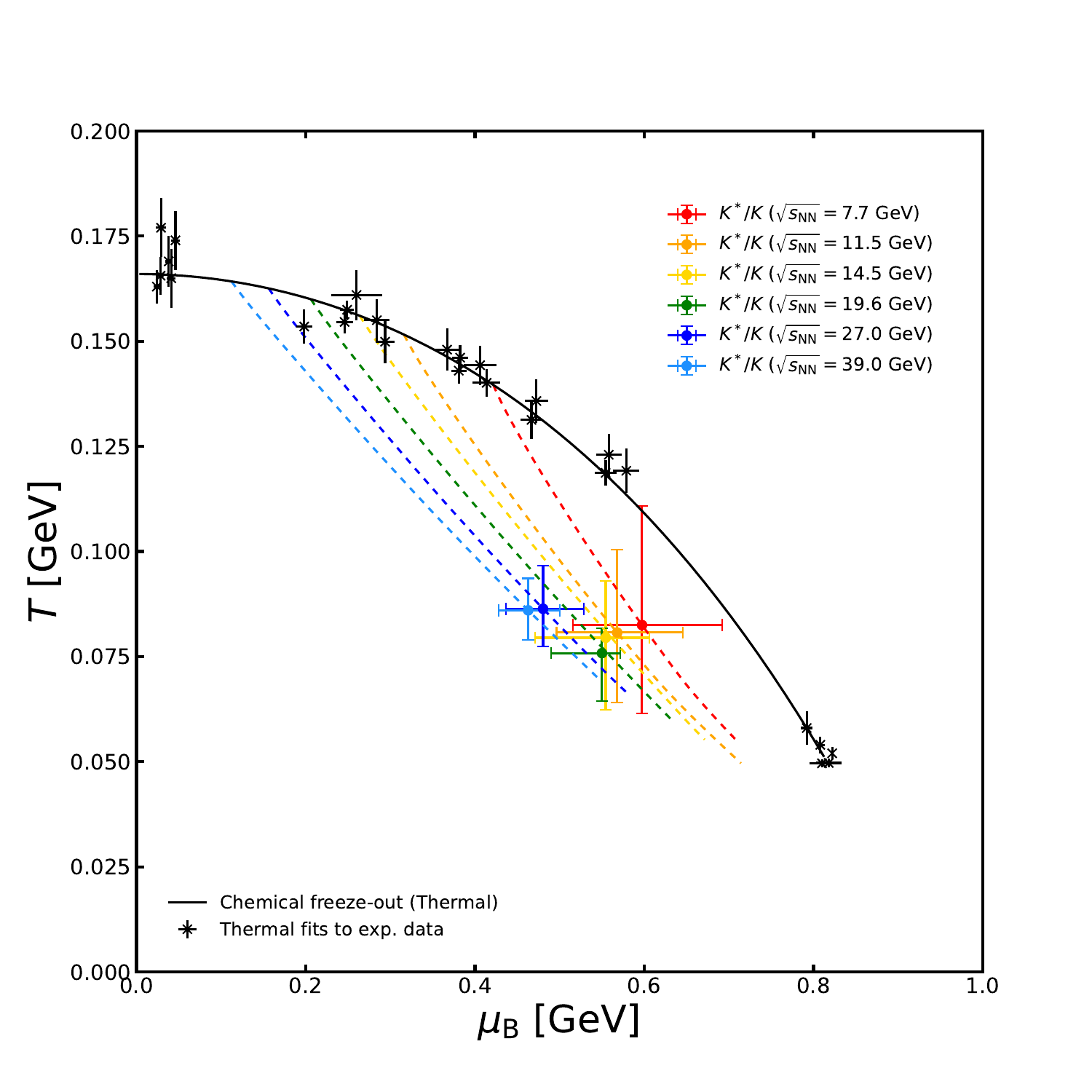}
    \caption{[Color online] Kinetic freeze-out points of the $K^*$ resonances (colored circles with error bars) and their trajectories from chemical to kinetic freeze-out (colored dashed lines) calculated with the rate equation network. Also shown are thermal fits to experimental data, respective the chemical freeze-out data, and their parametrization \cite{Cleymans:2006qe}.}
    \label{fig:phase_diagram}
\end{figure}

\begin{figure} [t!]
    \centering
    \includegraphics[width=\columnwidth]{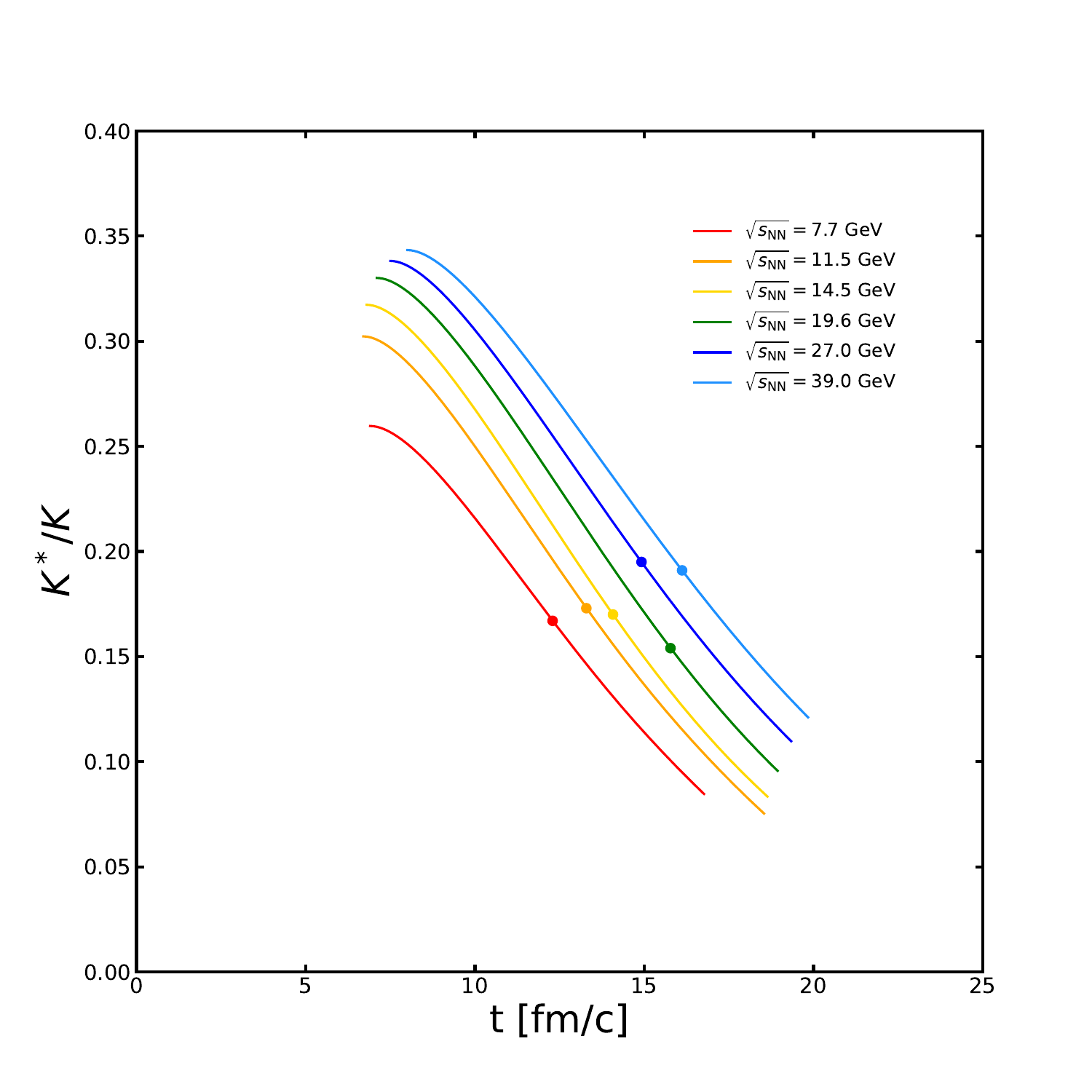}
    \caption{[Color online] Time evolution of the $K^*/K$ ratio starting from the chemical freeze-out time obtained from the full set of rate equations in central Au+Au reaction from $\sqrt{s_\mathrm{NN}}=7.7$ GeV to $\sqrt{s_\mathrm{NN}}=39$ GeV (denoted by differently colored lines). The time at which the ratio matches the measured $K^*/K$ value is denoted by a circle.}
    \label{fig:time_evol}
\end{figure}

\begin{figure} [t!]
    \centering
    \includegraphics[width=\columnwidth]{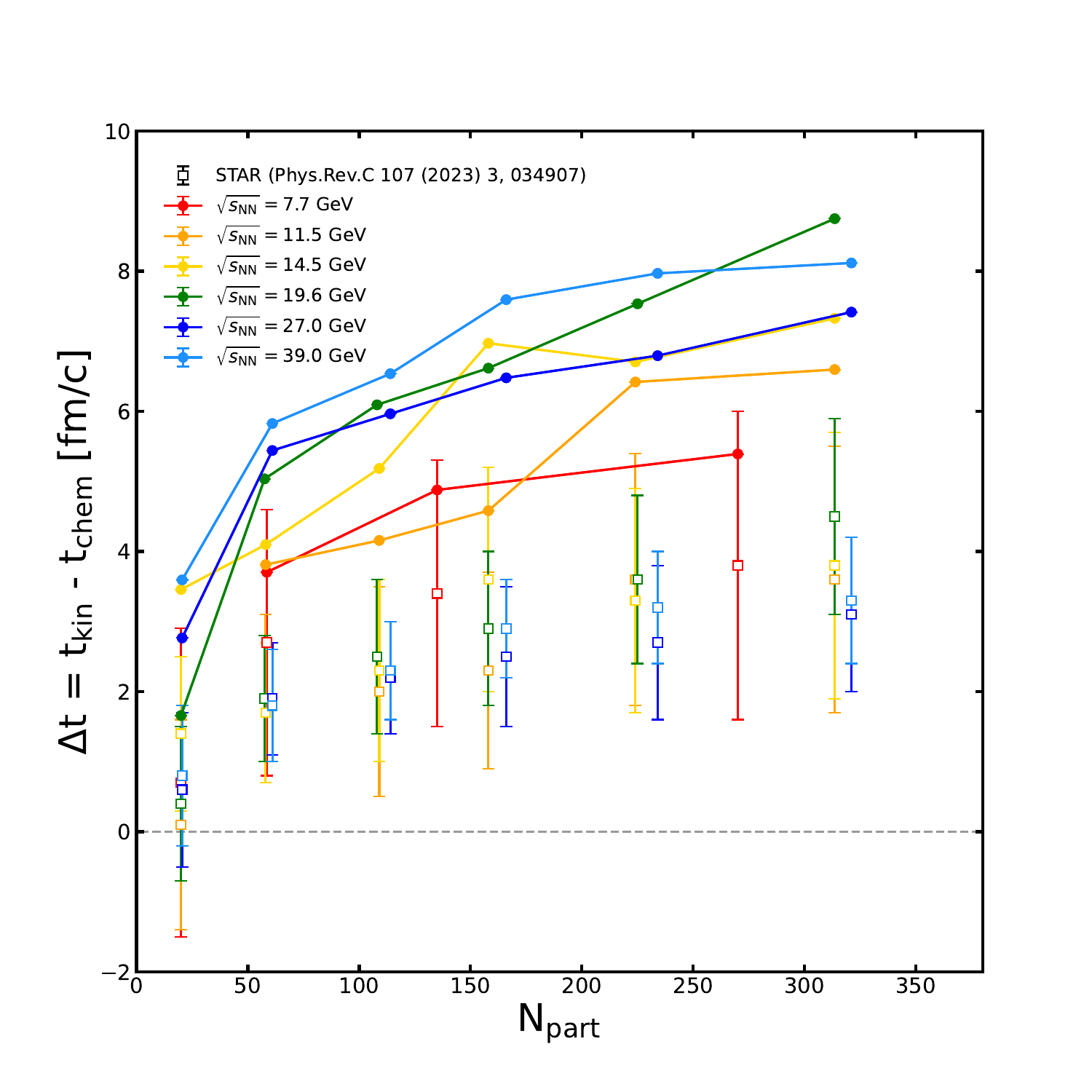}
    \caption{[Color online] Duration of the hadronic phase (i.e. $t_\mathrm{kin} - t_\mathrm{ch}$) in Au+Au reactions from $\sqrt{s_\mathrm{NN}}=7.7$ GeV to $\sqrt{s_\mathrm{NN}}=39$ GeV as function of centrality. The full circles with solid lines denote the calculation using the full set of rate equations including regeneration, while the open squares with error bars show the results obtained by STAR \cite{STAR:2022sir}, neglecting the regeneration processes. }
    \label{fig:time_npart}
\end{figure}

\section{Freeze-out}
Given the above described set-up, the rate equation network is run for Au+Au reactions for the range of collision energies recently investigated by the STAR collaboration, i.e. from $\sqrt{s_\mathrm{NN}}=7.7$ GeV to $\sqrt{s_\mathrm{NN}}=39$ GeV. In Fig. \ref{fig:Multiplicity_ratios_edep} we show the $K^*/K$ ratio in Au+Au reactions as a function of centrality for the different collision energies from $\sqrt{s_\mathrm{NN}}=7.7$ GeV (upper left) to $\sqrt{s_\mathrm{NN}}=39$ GeV (lower right). Full colored symbols with bands denote the rate equation network calculations, while the open black symbols show the experimental data from the STAR collaboration \cite{STAR:2022sir}. One observes that the rate equation network allows for a very good description of the measured data.

Next we analyze the freeze-out position of the $K^*$ resonances and the trajectories of the system evolution in the $T-\mu_B$ diagram. Fig. \ref{fig:phase_diagram} shows the kinetic freeze-out points of the $K^*$ resonances (colored circles with error bars) in the phase diagram. Also shown are the trajectories of the system (colored dashed lines) obtained from the rate equations, starting on the chemical freeze-out line and passing through the kinetic freeze-out points. The calculated freeze-out points of the $K^*$ resonances clearly fall below the chemical freeze-out points measured in different experiments (black symbols). 

\section{Time evolution}
After benchmarking the bulk properties, we focus now on the time evolution and the duration of the hadronic stage. Fig. \ref{fig:time_evol} shows the time evolution obtained from the full set of rate equations of the $K^*/K$ ratio in central Au+Au reaction from $\sqrt{s_\mathrm{NN}}=7.7$ GeV to $\sqrt{s_\mathrm{NN}}=39$ GeV. One generally observes that the time evolution is rather similar for all investigated energies. It is, however, of utmost importance that the time evolution is not just an exponential decay as assumed by the STAR collaboration (see Eq. \ref{eq:star}). In strong contrast, the time evolution of the ratio is essentially linear. This is due to the strong contribution of regeneration, i.e. $K+\pi\rightarrow K^*$ that counteracts the decay of the $K^*$. Such a behavior is of course expected by the principle of detailed balance and leads to a substantial correction. We find an (essentially) energy independent slope parameter $\mathrm{d} (K^*/K) / \mathrm{d} t \approx -0.02$ fm$^{-1}$. This provides an improved method to infer the life time of the hadronic phase from the experimentally measured $K^*/K$ ratio as follows:
\begin{align}
    \Delta t_\mathrm{hadronic} &= \frac{\left( K^*/K \right)_{\text{CFO}} - \left( K^*/K \right)_{\text{KFO}}}{0.02 \, \text{fm}^{-1}} .
\end{align}

We summarize our findings about the life time of the hadronic phase $\Delta t_\mathrm{hadronic}$ in Fig. \ref{fig:time_npart}. Here we show the duration of the hadronic phase in Au+Au reactions from $\sqrt{s_\mathrm{NN}}=7.7$ GeV to $\sqrt{s_\mathrm{NN}}=39$ GeV as a function of centrality. The full circles with solid lines denote the calculation using the full set of rate equations including regeneration, while the colored open squares with error bars show the results obtained by STAR \cite{STAR:2022sir}, neglecting the regeneration processes. One clearly observes that the inclusion of the regeneration process leads to a substantial increase in the estimated life time of the hadronic phase. The new estimates are also in line with previous estimates for the life time of the hadronic stage that also suggest a duration on the order of 4-8 fm/c \cite{Li:2007im,Graef:2012sh,Knospe:2015nva} for similar values of $\mathrm{d}N_{ch}/\mathrm{d}\eta$ and using  $\tau_\mathrm{emission}\approx \sqrt{R_o-R_s}/\beta$ to convert the HBT-radii to an emission duration \cite{Lisa:2016buz}.

\section{Conclusion}
We have analyzed the recent STAR data on the $K^*/K$ ratio measured in Au+Au reactions from $\sqrt{s_\mathrm{NN}}=7.7$ GeV to $\sqrt{s_\mathrm{NN}}=39$ GeV at various centralities. In contrast to the previously employed simplified estimates for the life time of the hadronic stage based only on the resonance decay time, we have shown that the results change drastically, if one uses the full set of coupled rate equations. The main difference is that the $K^*/K$ ratio does not decrease exponentially, but the regeneration leads to an essentially linear decrease over time. This in turn increases the estimated duration of the hadronic phase to meet the experimentally measured values. We show that the hadronic life time estimated here is a factor of 2-4 larger then the STAR's estimate and is now in line with other estimates. The results are thus highly valuable for the current investigation of the $K^*/K$ ratio conducted by the NA61 collaboration.

\section*{Acknowledgments}
We thank A. Knospe and C. Markert for fruitful discussion during the Strangeness in Quark Matter conference 2024 in Strasbourg. 
We further thank V. Vovchenko for inspiring discussions about partial chemical equilibrium.
T.R. acknowledges support through the Main-Campus-Doctus fellowship provided by the Stiftung Polytechnische Gesellschaft (SPTG) Frankfurt am Main and moreover thanks the Samson AG for their support.
The computational resources for this project were provided by the Center for Scientific Computing of the GU Frankfurt and the Goethe--HLR.
Also this research has received funding support from the NSRF via the Program Management Unit for Human Resources \& Institutional Development, Research and Innovation [grant number B16F640076].


\end{document}